\documentclass{article}
\usepackage{spconf,amsmath,graphicx}

\usepackage{graphicx}
\usepackage{multirow}
\usepackage{booktabs}
\usepackage{tabularx, color, soul}

\title{Versatile Video coding and super-resolution for efficient delivery of 8K video with 4K backward-compatibility}
\name{Charles Bonnineau$^{\star \dagger \ddagger}$, Wassim Hamidouche$^{\star \ddagger}$, Jean-Fran\c cois Travers$^\dagger$ and Olivier D{\'e}forges$^\ddagger$}
\address{$^\star$ IRT b$<>$com, Cesson-Sevigne, France, \\
$^\dagger$TDF, Cesson-Sevigne, France, \\
$^\ddagger$Univ Rennes, INSA Rennes, CNRS, IETR - UMR 6164, Rennes, France}

\begin{document}
\ninept
\maketitle
\begin{abstract}

In this paper, we propose, through an objective study, to compare and evaluate the performance of different coding approaches allowing the delivery of an 8K video signal with 4K backward-compatibility on broadcast networks. Presented approaches include simulcast of 8K and 4K single-layer signals encoded using High-Efficiency Video Coding (HEVC) and Versatile Video Coding (VVC) standards, spatial scalability using SHVC with 4K base layer (BL) and 8K enhancement-layer (EL), and super-resolution applied on 4K VVC signal after decoding to reach 8K resolution. For up-scaling, we selected the deep-learning-based super-resolution method called Super-Resolution with Feedback Network (SRFBN) and the Lanczos interpolation filter. We show that the deep-learning-based approach achieves visual quality gain over simulcast, especially on bit-rates lower than 30Mb/s with average gain of 0.77dB, 0.015, and 7.97 for PSNR, SSIM, and VMAF, respectively and outperforms the Lanczos filter in average by 29\% of BD-rate savings.  
\end{abstract}
\section{Introduction}
\label{sec:intro}


With the deployment of the latest Ultra High Definition Television (UHDTV) system \cite{itu-r_recommendation_BT2020-1}, it is projected to improve the Quality of Experience (QoE) through the introduction of new features to the existing High Definition Television (HDTV) system~\cite{itu-r_recommendation_BT709-5}, such as High Dynamic Range (HDR), wider color gamut, High Frame-Rate (HFR) and higher spatial resolutions including 4K (3840x2160) and 8K (7680x4320). The delivery of such video formats on current broadcast infrastructures is a real challenge and requires efficient compression methods to reach the available throughput while ensuring high video quality.

Recent studies on 8K contents coding~\cite{ichigaya2016required, sugito2018study} have shown that a bandwidth of approximately 80Mb/s is required to reach a significant visual quality improvement over 4K video format using the High Efficiency Video Coding (HEVC)~\cite{sullivan2012overview} standard. The Joint Video Exploration Team (JVET) established by the International Telecommunication Union (ITU-T) and Motion Picture Experts Group (MPEG) has investigated the development of the next generation video coding standard, called Versatile Video Coding (VVC). This standard, scheduled to be finalized in late 2020, should make the delivery of new video formats, including 8K, more affordable as it is projected to offer 30-50\% of bit-rate savings over HEVC~\cite{Sidaty_2019_PCS}. 

\begin{figure}[htb]

\begin{minipage}[b]{1.0\linewidth}
  \centering
  \centerline{\includegraphics[width=\linewidth]{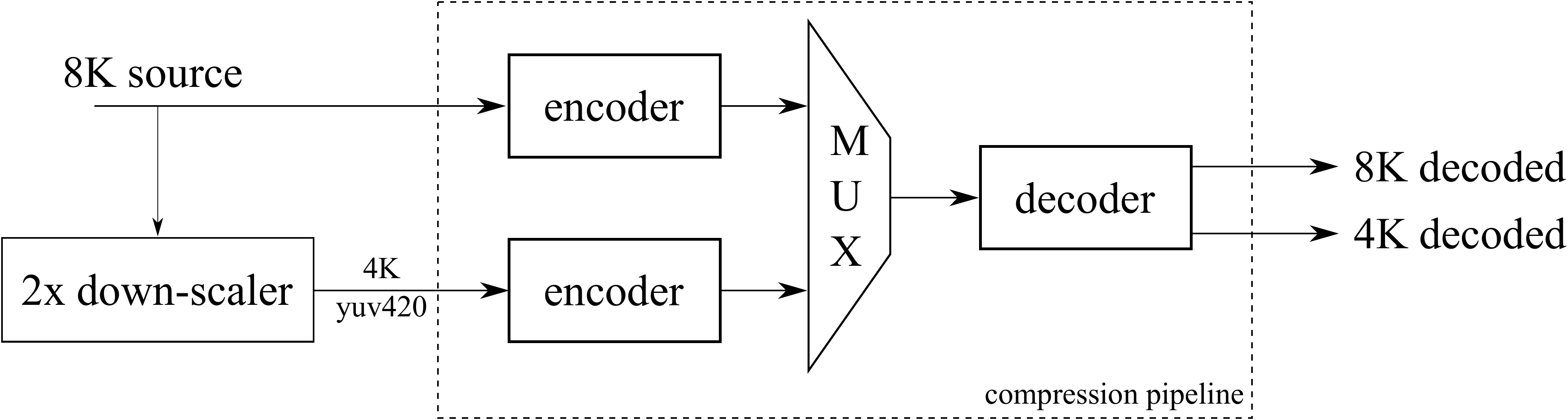}}
  \centerline{(a) Simulcast}\medskip
\end{minipage}

\begin{minipage}[b]{1.0\linewidth}
  \centering
  \centerline{\includegraphics[width=\linewidth]{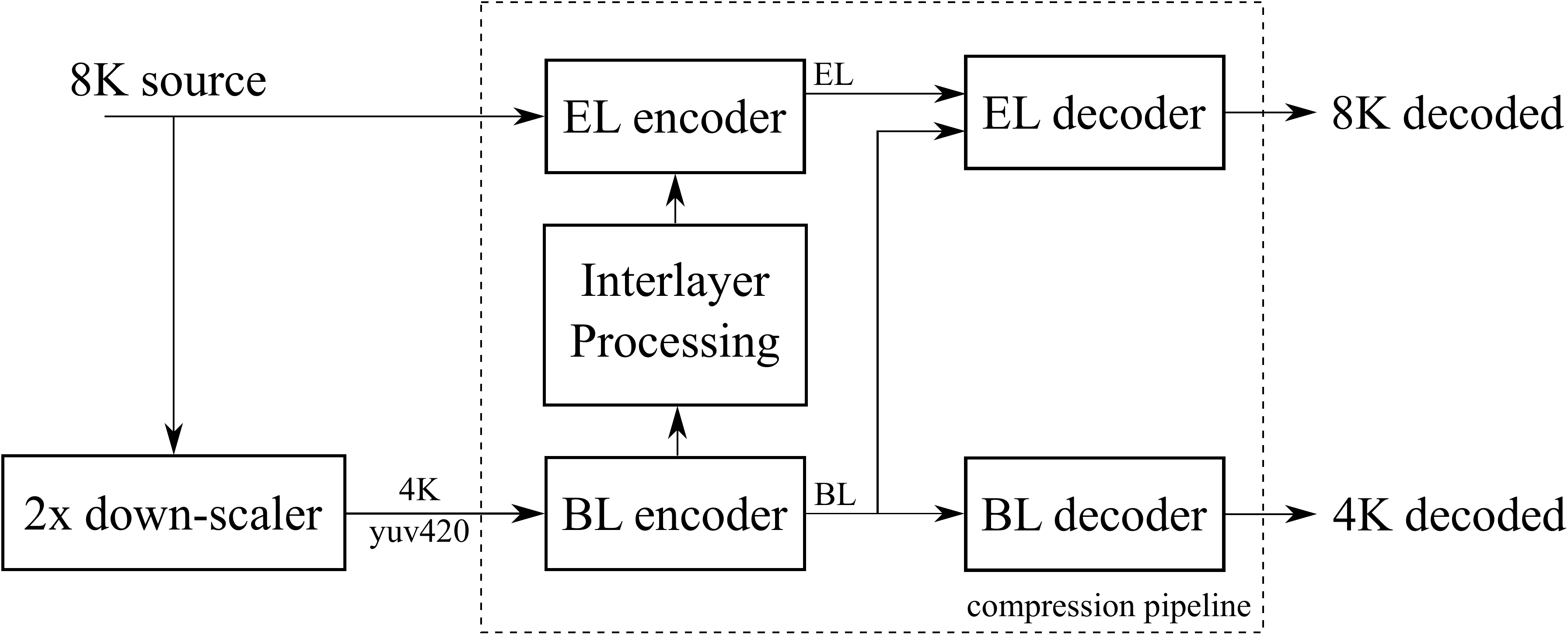}}
  \centerline{(b) Spatial scalability}\medskip
\end{minipage}
\hfill
\begin{minipage}[b]{1.0\linewidth}
  \centering
  \centerline{\includegraphics[width=\linewidth]{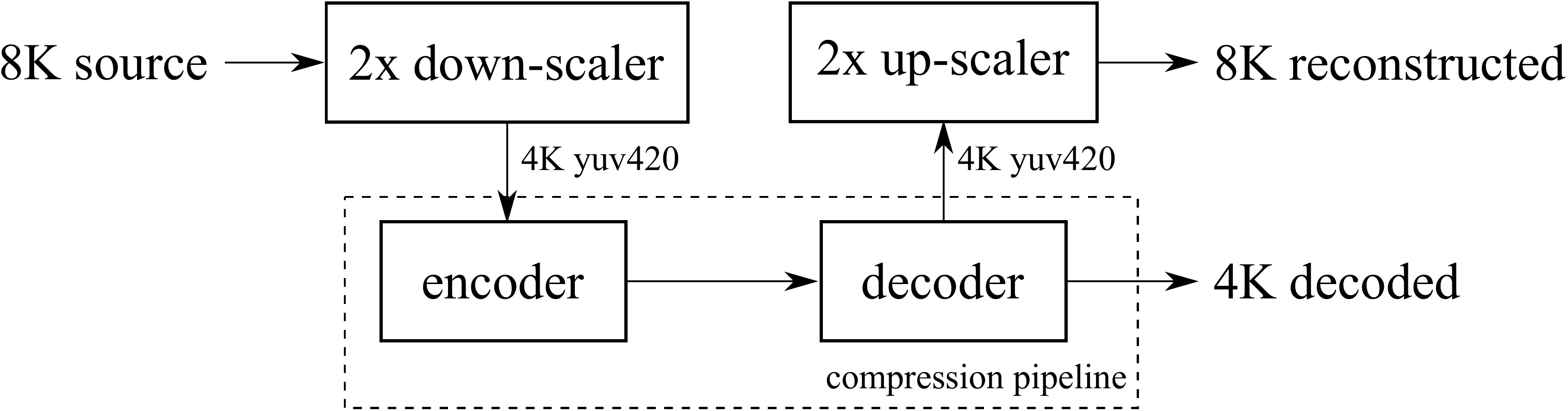}}
  \centerline{(c) Pre/post-processing}\medskip
\end{minipage}
\caption{Coding approaches that enable 8K signal to be delivered with 4K backward-compatibility.}
\label{fig:tested_configurations}
\end{figure}

On the other hand, to ensure service continuity, backward compatibility with UHD-1 signals shall be supported as a first step to provide both 4K and 8K resolutions. Simulcasting 4K and 8K single-layer signals would be the simplest solution as no additional constraint is added to the decoder. However, available bandwidth can vary depending on the exploited transmission network, reducing the number of use-cases that can be covered using this approach. For instance, in the context of terrestrial transmission, the bandwidth is limited in the range 30-40Mb/s using practical DVB-T2~\cite{etsi_dvb-t2} channels while satellite infrastructures allow a bandwidth up to 80Mb/s relaying on a complete DVB-S2X~\cite{etsi_dvb-s2x} transponder or multiple bonded transponders. Other coding approaches, such as spatial scalability using the scalable extension of HEVC called Scalable High Efficiency Video Coding (SHVC)~\cite{boyce2015overview}, can be considered for hybrid mechanism with base and enhancement layers to receive either 4K and 8K resolutions. To avoid using scalable-compliant decoders, another solution would require pre and post-processing steps into the classical transmission pipeline, as described in Figure~\ref{fig:tested_configurations}. With the progress in Deep Learning for image processing, learning-based spatial up-scalers \cite{dong2015image, shi2016real, ledig2017photo} have outperformed classical interpolation methods such as bicubic \cite{keys1981cubic} or Lanczos \cite{duchon1979lanczos} filters allowing a high resolution to be more accurately recovered from a lower resolution.

In this paper, we investigate the performance of these three approaches to transmit 8K video contents while ensuring backward compatibility with 4K devices. This study will focus on the performance of VVC for the simulcast and pre/post-processing approaches. As scalability is not yet integrated into the VVC test model, despite being planned to be released in the first version of the standard \cite{jvet-o0135}, we are also considering  HEVC and SHVC for simulcast and spatial-scalability to estimate its possible benefits in the considered context. For the experiments, the coding performance is assessed with objective quality metrics, including PSNR, SSIM and VMAF \cite{vmaf}. Although VMAF is optimized for visual quality estimation of 4K contents, it can be relevant to add this evaluation method in the experiment as it proposes a high correlation with subjective test ratings. For the pre/post-processing pipeline, we use the deep-learning-based method Super-Resolution with Feedback Network (SRFBN) \cite{Li_2019_CVPR}, which enables good performance in both visual quality improvement and runtime. We have trained the model using compressed data to propose a fair evaluation of this method on contents presenting compression artefacts. For VVC, this study shows that, for bit-rates lower than 30Mb/s, the tested super-resolution method offers an average gain over simulcast of 0.77dB, 0.015, and 7.97 regarding PSNR, SSIM, and VMAF, respectively. Moreover, this method outperforms the Lanczos filter by offering about 29\% of BD-rate savings in PSNR. Thus, this approach is particularly effective for the transmission of 8K contents with 4K backward compatibility using existing terrestrial infrastructures.     

The remainder of this paper is organized as follow. Section~\ref{sec:backward} presents the different tested coding schemes. Section~\ref{sec:experimental_settings} gives the video sequences and the test conditions. Results are then presented and analyzed in Section~\ref{sec:results}. Finally, Section~\ref{conclusion} concludes this paper. 

\section{Backward-compatible approaches for 8K video coding}
\label{sec:backward}

A mechanism for 8K transmission that enables backward compatibility with existing UHD-1 receivers might help to increase audience reach. Solutions illustrated in Figure~\ref{fig:tested_configurations}, namely simulcast, spatial-scalability, and pre/post-processing, allow the delivery of both 4K and 8K signals. This section aims at presenting these approaches by introducing tools that enable each pipeline to be set-up. First, simulcast of two single-layer signals approach is presented. Then, spatial-scalability using scalable-compliant codec such as SHVC is developed. Finally, we present a pre/post-processing pipeline that enables both resolutions to be provided from a unique 4K signal.

\subsection{Simulcast}
\label{subsec:simulcast}
Contributions to compression standards like HEVC or its successor VVC, scheduled to be released in 2020, enable video signal compression to be continuously improved through the MPEG standardisation body. Simulcast is the process of transmitting several versions of an input signal encoded with single-layer coding approaches such as HEVC or VVC to cover different target outputs. Although HEVC has brought a significant bit-rate reduction for 4K delivery, its efficiency is not sufficient for 8K applications. For instance, the studies conducted in ~\cite{ichigaya2016required, sugito2018study} have shown that the bit-rate required by HEVC for 8K applications in 60Hz and 120Hz (temporally scalable) is around 80Mbps. In practice, HEVC codecs have been used for satellite broadcasting in Japan, where HEVC codec for 8K 120Hz has been developed~\cite{sugito2018study, sugito2015hevc}. For satellite transmission with DVB-S2X, bandwidth in the range 70-80Mb/s can be reached with the use of a complete transponder or multiple bonded transponders. For terrestrial transmission, such bandwidth requirements prevent the deployment of 4K and 8K simultaneously, as practical DVB-T2 \cite{eizmendi2014dvb} channels offer bandwidth in the range 30-40Mb/s over an 8MHz channel.  

\subsection{Spatial scalability}
\label{subsec:spatial-scalability}

To increase the coding efficiency, one possibility is to take advantage of the existing correlations between 4K and 8K signals by using a spatially-scalable codec with a base and enhancement layer model. In the case of SHVC for spatial scalability, a base-layer (BL) signal (low resolution) encoded with HEVC is used as a reference by an inter-layer processing module to encode the enhancement-layer (EL) signal (high resolution). The EL signal is described by the use of additional High-Level Syntax (HLS) and needs a scalable-compliant decoder to be decoded.  
Several standardization bodies such as the Advanced Television Systems Committee (ATSC) \cite{ATSC} or the Digital Video Broadcasting consortium (DVB) \cite{dvb} consider SHVC as a candidate for solving compatibility issues brought by new formats introduction.
However, due to codec compatibility issues caused by a late integration of the HEVC scalable extension, spatial-scalability is not much present in the current broadcast ecosystem. To tackle these obstacles and make the deployment of this technology more likely in the future, scalability is planned to be integrated into the first release of the next generation video coding standard VVC.

\subsection{Pre and post-processing}
\label{subsec:post_proc}
To cover a wide range of compatible UHD-1 or UHD-2 receivers, one solution would be to apply down-scaling and up-scaling operations to the signal outside the coding pipeline. Thus, the bandwidth is limited to broadcast 4K only while both resolutions can be displayed by the receiver. In the image processing field, the process of estimating a high-resolution version of a low-resolution content is referred to as super-resolution. In the last past years, learning-based super-resolution approaches have outperformed state-of-the-art methods through the last progress in the Artificial Intelligence (AI) field. The objective of these methods is to learn the non-linearity that exists between low-resolution images (LR) and their high-resolution version (HR) by analysing local statistics. For our study, we have selected the Feedback Network for Super-Resolution (SRFBN) as the up-scaling operator, which provides good performance in both visual quality enhancement and runtime. This approach is based on an end-to-end Convolutional Neural Network (CNN) coupled with a feedback mechanism that aims to output the best high-resolution (HR) version of the input low-resolution (LR) content. This method is optimized to recover details from uncompressed LR images by iterative minimization of $L1$ loss between the reconstructed HR image and the corresponding HR ground truth to increase its accuracy over training steps. Initially, the data-sets used to train the network are the publicly-available super-resolution image data-sets Flickr2K and DIV2K \cite{Agustsson_2017_CVPR_Workshops}. However, the low-resolution sequences used for our study can present strong compression artefacts after decoding, making the baseline SRFBN network not adapted to the target task. Thus, to perform a fair evaluation, we have trained the model by using the initial data-sets DIV2K and Flickr2K encoded with VTM in All-Intra coding configuration. More details on the training process are provided in Section~\ref{subsubsec:Super_resolution_settings}.    

\section{Experimental settings}
\label{sec:experimental_settings}

This section gives details of the experiment. First, the sequences selected for the study and their specifications are presented. Then, the codec specifications and super-resolution settings are detailed.

\subsection{Test sequences}
\label{subsec:test_sequences}

\begin{figure}[t]
  \includegraphics[width=\linewidth]{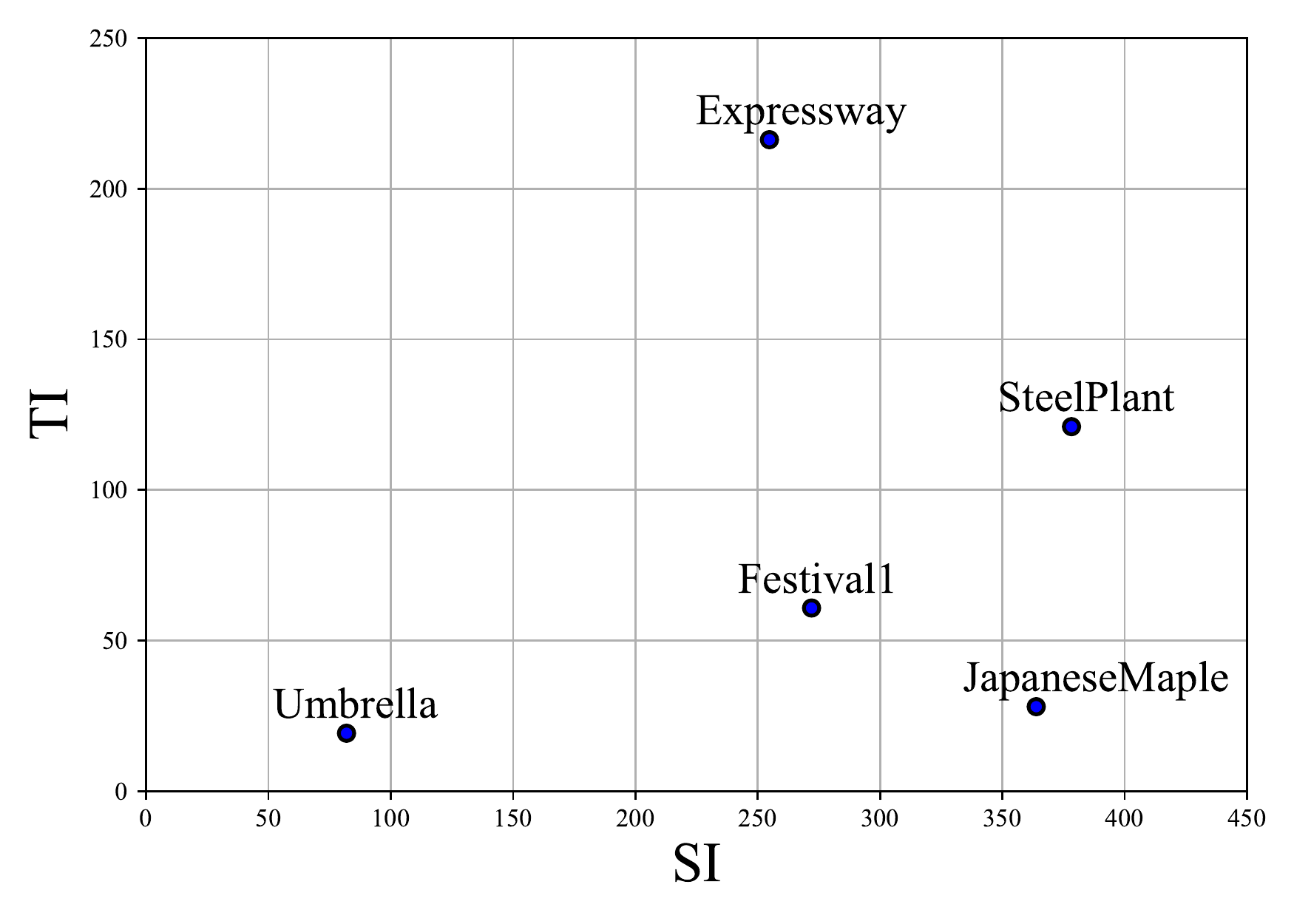}
  \caption{SI-TI graph of the tested video sequences.}
  \label{fig:SITI}
\end{figure}

For the experiment, we have selected five 8K sequences of 5 seconds. These sequences were provided by The Institute of Image Information and Television Engineers. The spatial and temporal information (SI-TI) ~\cite{itu-r_recommendation_BT910} of these sequences is plotted in Figure~\ref{fig:SITI}. This 2D-plan shows that the contents selected for the study have various spatial and temporal features. The objective is to have a diversity of contents and analyse the coding efficiency regarding their features. The sequences in 4K resolution are generated by a bicubic down-scale and the SHM down-scaling filter described in \cite{boyce2015overview} for the pre/post-processing and spatial-scalability approaches, respectively.  


\subsection{Test conditions}
\label{subsec:test_conditions}

\subsubsection{Verification models}
\label{subsubsec:verification_models}

\begin{table}[!t]
\renewcommand{\arraystretch}{1.3}
\small
\caption{Standard verification models specification}
\label{vm_spec}
\centering
\begin{tabular}{l|ccc}
\midrule[0.3mm]
\bfseries Standard & \bfseries VVC & \bfseries HEVC & \bfseries SHVC\\
\midrule[0.3mm]
Reference Softw. & VTM-5.0 & SHM-9.0 (BL) & SHM-9.0 \\
Profile & next & main10 & main10 \\
GOP Size & 16 & 16 & 16 \\
\midrule[0.3mm]
\end{tabular}
\end{table}

Verification software models provide a reference implementation of different compression standards representing their maximal performance offered with a moderate level of optimization. 
For the experiment, the Common Test Conditions for VTM \cite{reference_software_vvc} and SHM \cite{reference_software_shvc} are used to provide a fair rate/distortion estimation. HEVC simulcast is simulated using the SHM base-layer mode to ensure a fair comparison over spatial-scalability. The coding configurations are summarized in Table~\ref{vm_spec} for both codecs. Each test sequence is encoded in Random Access coding configuration with Quantization Parameter (QP) values of 22, 27, 32, and 37. QP of 17 and 42 are added to the parameter-set for the pre/post-processing and simulcast evaluations, respectively, to cover a similar range of bit-rate. To assess the visual quality, objective metrics, including PSNR-Y, SSIM, and VMAF, are used to measure the distortion between the decoded or reconstructed 8K signal and the original one. To quantify the average gain in bit-rate or visual quality offered by a tested method over another, we use the Bj\o ntegaard delta (BD) method described in \cite{vceg_m33}. For pre/post-processing, the bit-rate is assessed on the 4K single-layer signal. For simulcast and spatial-scalability, the measured bit-rate corresponds to the sum of both 8K and 4K bitstreams. 

\subsubsection{Super-resolution settings}
\label{subsubsec:Super_resolution_settings}

 The baseline version of SRFBN provided by the author in \cite{Li_2019_CVPR} is trained to recover the high-resolution version of uncompressed low-resolution data. In our case, we focus on assessing this method on video presenting compression artefacts. Learning-based super-resolution being very sensitive to the nature of the training data, a fair evaluation of this approach is not possible with the provided learned parameters. To propose a fair evaluation of the model, we have fine-tuned the baseline network using a compressed version of the initial image datasets DIV2K and Flickr2K. First, we have generated pairs of LR/HR images by applying a bicubic down-scaling filter with a scale-factor of 2 from HR images. Then, each LR image has been encoded using the All Intra configuration of the VTM with five QP values, including 17, 22, 27, 32 and 37 to cover a large panel of distortions. These coded samples, representing different levels of compression distortions, are used to fine-tune the network. To evaluate the performance of SRFBN, we have also compared results obtained with a Lanczos filter applied on the low-resolution sequences.

\section{Results}
\label{sec:results}

In this section, we present and analyze the results of the simulations conducted using the verification models and the settings described in Section~\ref{sec:experimental_settings} for the three approaches presented in Section~\ref{sec:backward}. First, we compare simulcast and spatial scalability using SHM. Then, we evaluate the pre/post-processing pipeline over simulcast with VTM.  

\subsection{Simulcast and spatial scalability}

\begin{table}[t]
	\vspace{-1em}
	\setlength{\tabcolsep}{12pt}
	\footnotesize
	\begin{center}
		\caption{BD-rate (\%) for SHVC compared to HEVC simulcast}
		\label{tab:bdrate_matrix}
		\begin{tabularx}{\linewidth}{l|ccc}
			\midrule[0.3mm]
			\bfseries Sequence & \bfseries PSNR-Y & \bfseries SSIM & \bfseries VMAF   \\
			\midrule[0.3mm]
		    Expressway & -8.86 & -8.23 & -8.47  \\
		    Festival1 & -12.97 & -12.07 & -12.19  \\
		    JapaneseMaple & -12.93 & -11.65 & -13.42  \\
		    SteelPlant & -16.40 & -14.86 & -12.66  \\
		    Umbrella & -17.88 & -18.73 & -14.13  \\
			\midrule[0.2mm]
			\bfseries Average & \bfseries -13.81 & \bfseries -13.11 & \bfseries -12.17\\
			\midrule[0.3mm]
		\end{tabularx}
	\end{center}
	\vspace{-2.5em}
\end{table}

\begin{figure*}[t]
\begin{minipage}[b]{0.33\linewidth}
  \centering
  \centerline{\includegraphics[width=1\linewidth]{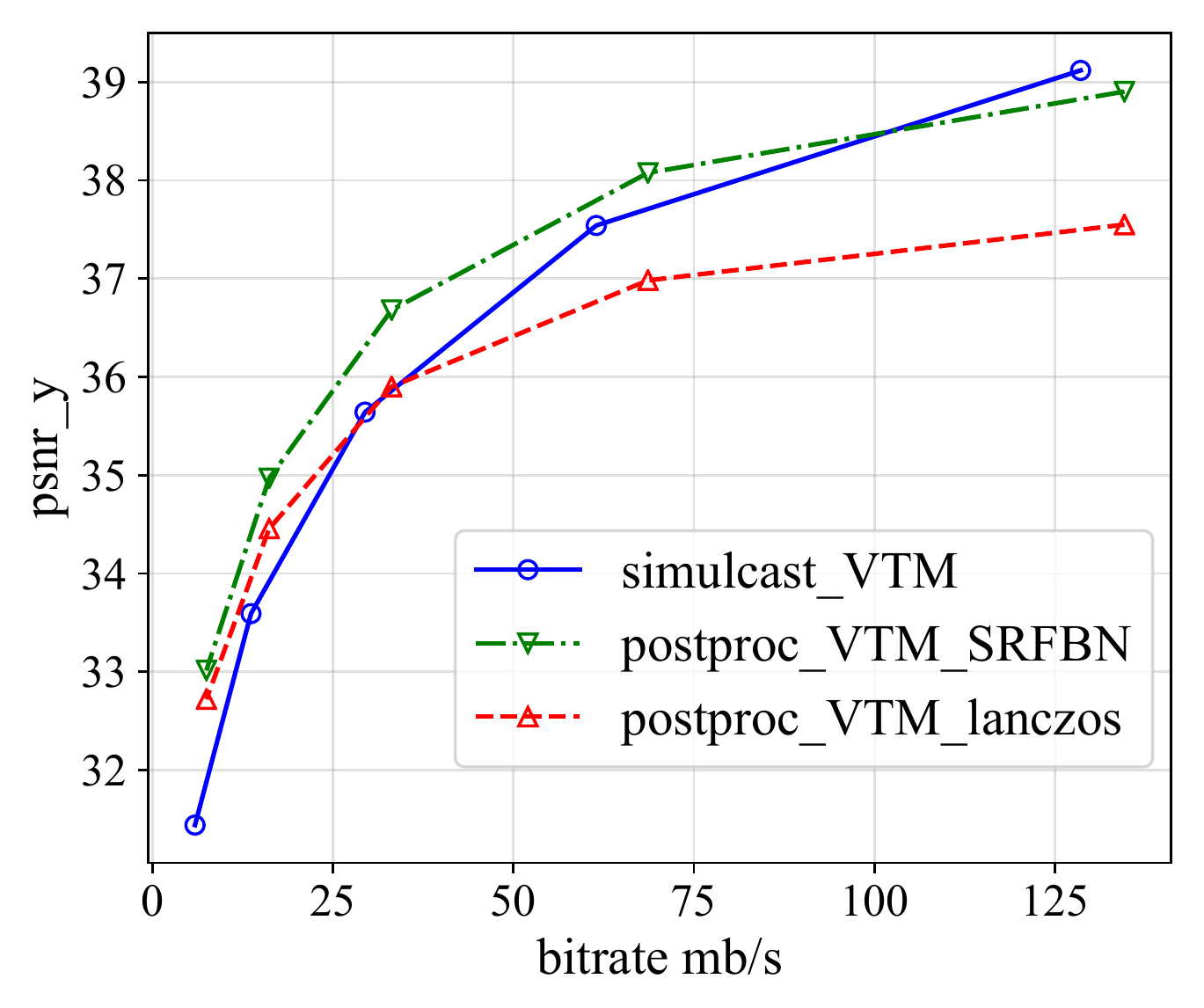}}
  \centerline{(a) PSNR-Y}\medskip
\end{minipage}
\hfill
\begin{minipage}[b]{0.33\linewidth}
  \centering
  \centerline{\includegraphics[width=1\linewidth]{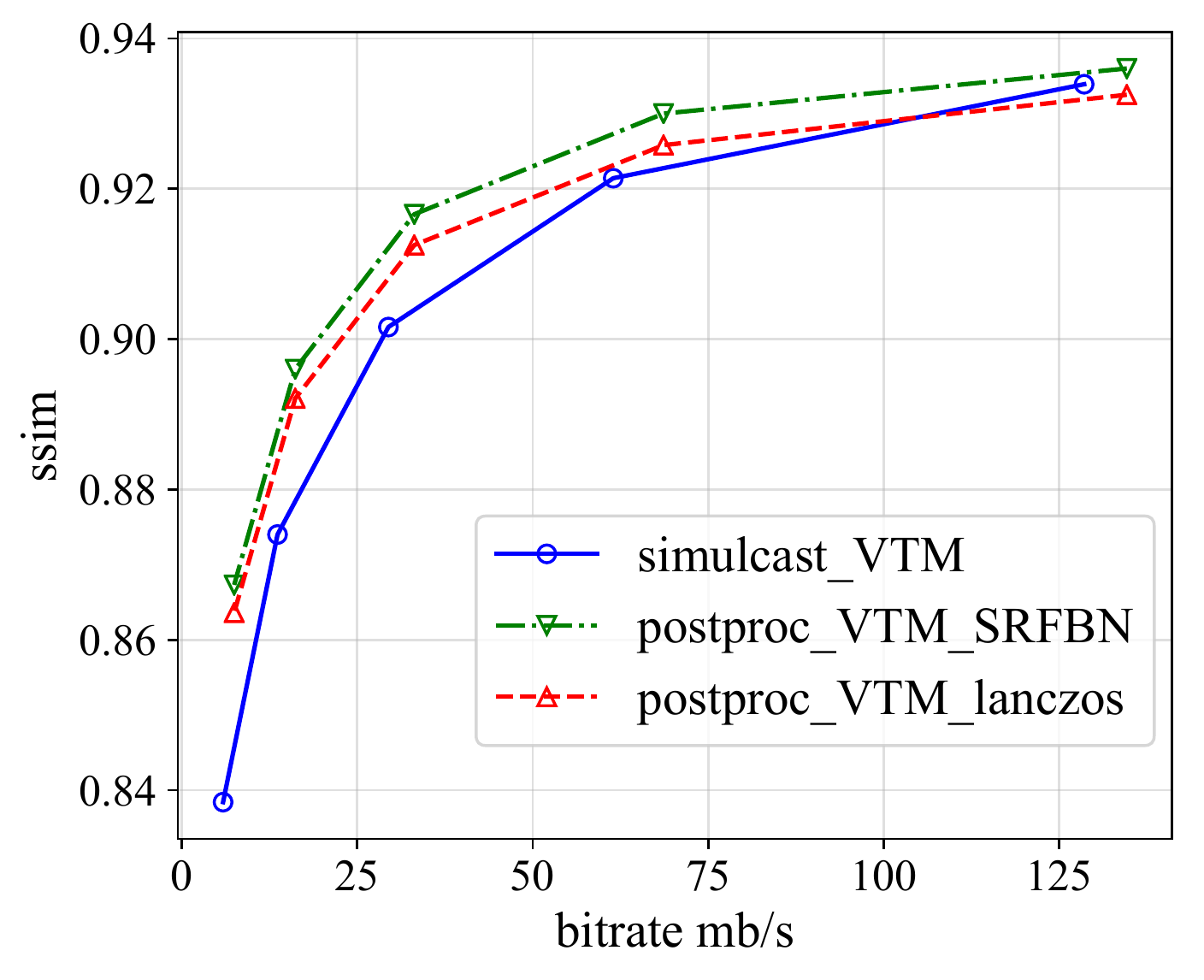}}
  \centerline{(b) SSIM}\medskip
\end{minipage}
\hfill
\begin{minipage}[b]{0.33\linewidth}
  \centering
  \centerline{\includegraphics[width=1\linewidth]{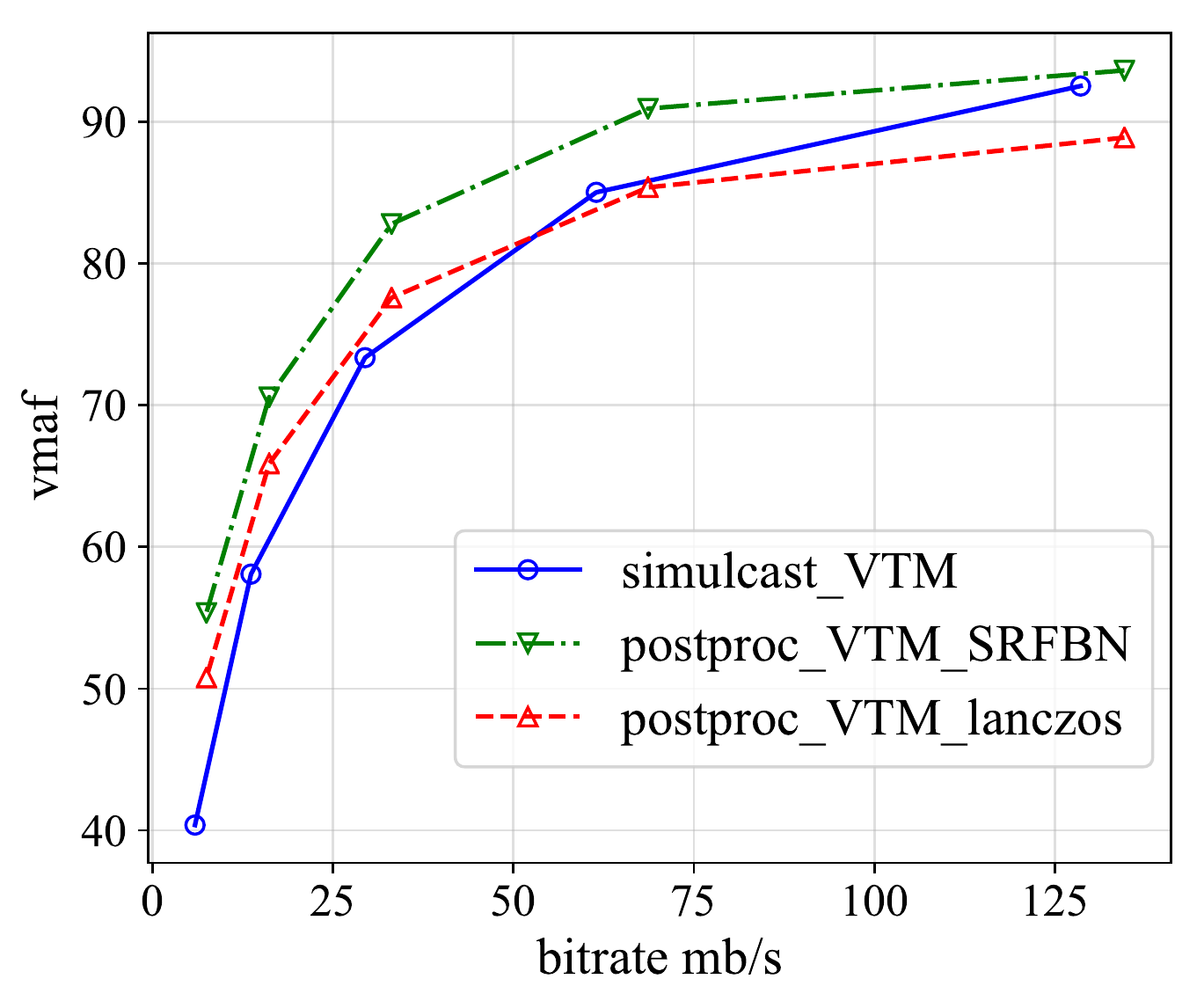}}
  \centerline{(c) VMAF}\medskip
\end{minipage}

\caption{Average rate-distortion curves on the five 8K video sequences of simulcast and pre/post-processing approaches.}
\label{fig:rate_dist}
\end{figure*}

One objective of this study is to estimate the potential gain of spatial scalability over simulcast of 8K and 4K. Since VVC does not yet support spatial scalability, we compare the simulcast of 4K and 8K in HEVC and SHVC with 4K base-layer and 8K enhancement-layer as a first step. Table~\ref{tab:bdrate_matrix} gives the BD-rate results of SHVC with respect to the HEVC simulcast configuration for the five video sequences. Inter-layer predictions enable 13.81\%, 13.11\%, and 12.17\% of average BD-rate savings over simulcast for PSNR, SSIM, and VMAF, respectively. However, scalable-compliant decoders will still be required to decode the SHVC bitstream.

\subsection{Pre and post-processing}

\begin{table}[t]
	\vspace{-1em}
	\scriptsize
	\begin{center}
		\caption{BD-metric results per bit-rate interval (Mb/s) for pre/post-processing with VVC compared to simulcast with VVC.}
		\label{tab:perf_bitrate_interval}
		\begin{tabularx}{\linewidth}{l|l|XXX|XXX}
			\midrule[0.3mm]
			&&\multicolumn{3}{c}{\textbf{Lanczos}} & \multicolumn{3}{c}{\textbf{SRFBN}}\\[0.4em]
			 Sequence & Interval &	BD-PSNR & BD-SSIM &	BD-VMAF	& BD-PSNR & BD-SSIM & BD-VMAF\\
			\midrule[0.3mm]
			\multirow{4}{*}{Expressway}&-30 & -1.67 & 0.001 & -3.01	& -0.12	& 0.004 & 3.10 \\
			&30-80 & - & - & -	& -	& - & - \\
			&+80 & - & - & -	& -	& - & - \\
			\midrule[0.2mm]
			\multirow{4}{*}{Festival1}&-30 & 0.92 & 0.018 & 6.44 & 1.42 & 0.022 & 10.63\\
			&30-80 & -0.41 & 0.005 & 1.03 & 0.6 & 0.009 & 5.59\\
			&+80 & -1.44 & 0 & -2.53 & -0.11 & 0.003 & 1.74\\
			\midrule[0.2mm]
			\multirow{4}{*}{JapaneseMpl}&-30 & 0.79 & 0.029 & 5.01	& 1.09	& 0.036 & 9.19 \\
			&30-80 & 0.61 & 0.02 & 4.96 & 1.19	& 0.027 & 9.61  \\
			&+80 & -0.45 & 0.004 & 1.54	& 0.70	& 0.012 & 6.76  \\
			\midrule[0.2mm]
			\multirow{4}{*}{SteelPlant}&-30& 0.52 & 0.007 & 3.34 & 1.01	& 0.011 & 10.65  \\
			&30-80 & 0.29 & 0.007 & 2.38 & 1.14	& 0.012 & 10.78 \\
			&+80 & -0.67 & 0.003 & -1.75 & 0.59	& 0.009 & 6.68  \\
			\midrule[0.2mm]
			\multirow{4}{*}{Umbrella}&-30 & 0.42 & 0.003 & 4.17	& 0.47	& 0.004 & 6.29  \\
			&30-80 & - & - & - & -	& - & -  \\
			&+80 & - & - & - & -	& - & - \\
			\midrule[0.3mm]
			\multirow{4}{*}{\textbf{Average}}&\textbf{-30} & \textbf{0.2} & \textbf{0.011} & \textbf{3.19}	& \textbf{0.77}	& \textbf{0.015} & \textbf{7.97}  \\
			&\textbf{30-80} & \textbf{0.1} & \textbf{0.006} & \textbf{1.67}	&\textbf{ 0.59}	&\textbf{0.01} & \textbf{5.2}  \\
			&\textbf{+80} & \textbf{-0.51} & \textbf{0.001} & \textbf{-0.55}	& \textbf{0.24}	& \textbf{0.005} & \textbf{3.04} \\[0.3em]
			\midrule[0.3mm]
		\end{tabularx}
	\end{center}
	\vspace{-2.5em}
\end{table}

In this section, we analyse the benefit of using a pre/post-processing pipeline instead of a simulcast approach for the delivery of 8K with 4K backward compatibility using VVC. First, a rate-distortion evaluation has been conducted using objective visual quality metrics, including PSNR, SSIM, and VMAF. The results for the average performance over all the tested sequences are illustrated in Figure~\ref{fig:rate_dist}. For all considered metrics, SRFBN enables average gain over Lanczos up to 29.02\%, 14.40\%, and 28.53\% of BD-rate savings regarding PSNR, SSIM and VMAF metrics, respectively. It can be noticed that the performance gap between SRFBN and Lanczos increases proportionally with the bit-rate increase. Although SRFBN is trained on compressed data, rough training methodology, as described in Section~\ref{sec:experimental_settings}, cannot allow the model to accurately recover details from LR sequences comprising strong compression artefacts. For broadcast, such artefacts can appear when the allocated bit-rate is not sufficient. To tackle this, it is possible to increase the efficiency for this use case by considering advanced training methods \cite{timofte2015seven} and/or adapted architecture to optimize the model for the target task.


The pre/post-processing pipeline is more efficient than simulcast until high bit-rate values (about 100Mb/s and 125Mb/s for SRFBN regarding PSNR and SSIM/VMAF, respectively). Regarding PSNR, SRFBN and Lanczos have better performance over simulcast until approximately 100Mb/s and 30Mb/s, respectively. Regarding SSIM and VMAF, SRFBN is more efficient than simulcast for all the presented bit-rate range. For VMAF, it can be explained by the use of detail loss metrics, namely, DLM \cite{li2011image} and VIF \cite{sheikh2006image}, in the final score computation. To evaluate the performance of SRFBN and Lanczos according to the bit-rate, we have computed the average visual quality gain for each metric over simulcast per tested sequence by using the BD-rate evaluation method.  Three bit-rate intervals are considered: lower than 30Mb/s, from 30Mb/s to 80Mb/s, more than 80Mb/s. Results are presented in Table~\ref{tab:perf_bitrate_interval}. We have computed additional QP points for some sequences to cover at least four points per bit-rate range. We can notice that the performance gap between SRFBN and Lanczos is more significant for the most complex sequences. Although being classified as a non-trivial sequence by the SITI graph, no results are collected after 30Mb/s for Expressway due to global motion, making the scene easy to predict by the encoder.
\section{Conclusion}
\label{conclusion}

Three approaches allowing the transmission of both 8K and 4K over broadcast networks have been assessed, including simulcast, spatial-scalability, and pre/post-processing. Waiting for more details on VVC scalable mode, a preliminary test using HEVC and its scalable extension SHVC was conducted. Experimental results have shown that spatial-scalability achieves 13.81\% of BD-rate savings compared to simulcast regarding PSNR. 

We have also demonstrated by a fair evaluation that the tested super-resolution method allows a bit-rate reduction of approximately 29\% in PSNR compared to Lanczos. This experiment also confirms that up-scaling 4K signal after decoding outperforms 4K and 8K simulcast using VVC, especially at bit-rates lower than 30Mb/s. Indeed, the BD evaluation demonstrates an average visual quality improvement over VVC simulcast of 0.77dB, 0.015, and 7.97 for PSNR, SSIM, and VMAF, respectively, on this bit-rate interval. Terrestrial broadcast limitations being in this range of bit-rate, the pre and post-processing approach may be preferred. However, the complexity of deep-learning-based tools is to be considered as it is added at the decoder-side. Future works will consider the VVC spatial scalability and subjective evaluation to consolidate the objective results. 

\newpage 
\bibliographystyle{IEEEbib}
\bibliography{IEEEexample}

\end{document}